\documentclass[3p]{elsarticle}
\usepackage{color}
\usepackage{ulem}
\usepackage{amsmath}
\usepackage{amssymb}
\usepackage{enumerate}
\usepackage{graphicx}
\usepackage{dcolumn}
\usepackage{bm}

\usepackage{hyperref}

\newtheorem{prop}{Proposition}[section]

\journal{Annals of Physics, \textbf{379} (2017) 13-33}

\begin{document}

\begin{frontmatter}

\title{Quantum mechanics of a constrained particle and the problem of prescribed
geometry-induced potential}

\author{Luiz C. B. da Silva$^*$}
\address{Departamento de Matem\'atica, Universidade Federal de Pernambuco, 50670-901, Recife, Pernambuco, Brazil}
\ead{luizsilva@dmat.ufpe.br}
\cortext[mycorrespondingauthor]{Corresponding author}

\author{Cristiano C. Bastos}
\address{Departamento de Qu\'imica, Universidade Federal Rural de Pernambuco, 52191-900, Recife, Pernambuco, Brazil}

\author{F\'abio G. Ribeiro}
\address{Laborat\'orio de F\'isica, Instituto Federal de Educa\c{c}\~ao, Ci\^encia e Tecnologia da Para\'iba - Campus Picu\'i, 58187-000, Picu\'i, Para\'iba, Brazil}

\begin{abstract}
The experimental techniques have evolved to a stage where various examples of nanostructures with non-trivial shapes have been synthesized, turning the dynamics of a constrained particle and the link with geometry into a realistic and important topic of research. Some decades ago, a formalism to deduce a meaningful Hamiltonian for the confinement was devised, showing that a geometry-induced potential (GIP) acts upon the dynamics. In this work we study the problem of prescribed GIP for curves and surfaces in Euclidean space $\mathbb{R}^3$, i.e., how to find a curved region with a potential given {\it a priori}. The problem for curves is easily solved by integrating Frenet equations, while the problem for surfaces involves a non-linear 2nd order partial differential equation (PDE). Here, we explore the GIP for surfaces invariant by a 1-parameter group of isometries of $\mathbb{R}^3$, which turns the PDE into an ordinary differential equation (ODE) and leads to cylindrical, revolution, and helicoidal surfaces. Helicoidal surfaces are particularly important, since they are natural candidates to establish a link between chirality and the GIP. Finally, for the family of helicoidal minimal surfaces, we prove the existence of geometry-induced bound and localized states and the possibility of controlling the change in the distribution of the probability density when the surface is subjected to an extra charge.
\end{abstract}


\end{frontmatter}


\section{\label{sec:Intro}Introduction}

The study of new material properties due to its shape has gained importance since the birth of nanoscience. The experimental techniques have evolved to a stage where various examples of nanostructures whose shape resembles planes, spheres, cylinders, and other non-trivial geometries, have been synthesized \cite{TerronesNewJPhys,CastroNetoRepProgPhys}. However, it is still difficult to establish a relation between the geometry and the quantum behavior of such systems. In face of these developments, writing the appropriate equations for a particle confined somewhere is essential to a proper understanding and modeling of these phenomena. In the 1950s De Witt addressed the problem of describing a confinement in a curved space through a quantization procedure, which resulted however in an ordering ambiguity \cite{DeWittRMP1957}. Later on, an approach which does not suffer from this ambiguity was devised by Jensen and Koppe \cite{JensenKoppeAnnPhys} in the 1970s and by Da Costa \cite{DaCostaPRA1981,DaCostaPRA1982} in the 1980s, showing that a geometry-induced potential (GIP) acts upon the dynamics \footnote{For more rigorous studies, see e.g. \cite{FroeseCMP2001,MitchellPRA2001,SchusterJaffeAnnPhys2003}; for studies allowing the confinement to vary along the constraint region, see e.g. \cite{WachsmuthPRA2010,SchmelcherPRA2014}, in which case analogies with the Born-Oppenheimer approximation is an important tool \cite{JeckoJMP2014}.}. Since then, some research on the subject has been reported, such as a path integral formulation \cite{MatsutaniJPhysSocJapan,MatsutaniPRa}, effects on the eigenstates of nanostructures \cite{EncinosaPRA,GravesenPRA}, interaction with an electromagnetic potential \cite{IkegamiSurfScience,FerrariPRL,deOliveiraJMathPhys,BrasileirosAnnPhys}, modeling of bound states on conical surfaces \cite{MoraesAnnPhys,BrasileirosJMP,ChinesesPhysicaE}, spin-orbit interaction \cite{EntinPRB,OrtixSpin,OrtixPRB2015}, electronic transport in nanotubes \cite{FernandosEfranceses,MarchiEtAlPRB2005}, and bent waveguides \cite{SchmelcherPRA2014,KrejcirikJGP2003,delCampoSciRep2014,HaagAHP2015}, just to name a few. 

For surfaces in $\mathbb{R}^3$, Encinosa and Etemadi \cite{EncinosaPRA} found that the shift in the ground-state energy may be of sufficient order to be observable in quantum nanostructures. More recently, on the experimental side, Onoe {\it et al.} reported on the observation of Riemannian geometric effects through the GIP on the Tomanaga-Luttinger liquid exponent in a 1D metallic $C_{60}$ polymer with an uneven periodic peanut-shaped structure \cite{OnoePRB,OnoeEPL}. In addition, Szameit {\it et al.} described the experimental realization of an optical analogue of the GIP \cite{SzameitEtAlPRL} (for more on the interplay between geometry and optics, see e.g. \cite{SchultheissPRL2010,PedersenJMO2015}). 

Inverse problems constitute an important subject from both  experimental and theoretical viewpoints, a classical problem being that of hearing the shape of a drum \cite{KacMonthly1966}, i.e., the determination of information about the geometry of a region that gives rise to a prescribed spectrum. More recently, we can mention the success in the detection of gravitational waves \cite{AbbottPRL2016}, which allows one to infer information about the spacetime geometry via measurements of an interferometric gravitational-wave detector. In the context of the constrained quantum dynamics this kind of problem offers the possibility of engineering surfaces and curves with a quantum behavior prescribed {\it a priori} through their geometry-induced potential and has been already investigated for curves \cite{delCampoSciRep2014} and a class of revolution surfaces \cite{AtanasovPLA2007}. Nonetheless, a comprehensive understanding of such an inverse problem for the constrained dynamics is still absent.

In this work we address the problem of prescribed GIP for curves and for surfaces in Euclidean space $\mathbb{R}^3$. The former can be easily solved by integrating the Frenet equations, while the latter involves the solution of a non-linear 2nd order PDE. We restrict ourselves to the study of surfaces invariant by a 1-parameter group of isometries of $\mathbb{R}^3$, which turns the PDE for the prescribed GIP into an ODE and leads us to the study of cylindrical, revolution, and helicoidal surfaces. The latter class is particularly important due to the fact that, by screw-rotating a curve clock- and counterclockwisely, one can generate pairs of enantiomorphic surfaces, which turn these objects the natural candidates to test and exploit a link between chirality and the effects of a GIP. We show how to find helicoidal surfaces associated with a given non-negative function and further specialize to the study of helicoidal minimal surfaces. For this class of minimal surfaces we prove the existence of localized and geometry-induced bound states, then generalizing known results for the dynamics on a helicoid \cite{AtanasovPRB2009}, and also the possibility of controlling the change in the distribution of the probability density when the surface is subjected to an extra charge.

The remaining of this work is organized as follows. In Section 2 we introduce notations, background material, and discuss the confining formalism and appearance of a GIP. In Section 3 we address the prescribed GIP problem and introduce the notion of invariant surfaces. Sections 4, 5, and 6, are devoted to the study of the geometry of cylindrical, revolution, and helicoidal surfaces, respectively. Further, in Section 7, we discuss some possible physical effects connected with the geometric properties of invariant surfaces, with a special emphasis on the dynamics on helicoidal minimal surfaces. Finally, in Section 8 we draw some conclusions.

\section{Preliminaries}

\subsection{Differential geometry background}

The Schr\"odinger equation on a Riemannian manifold $(M^n,g)$ for a particle of mass $m$ under the influence of a potential $V$ is written as
\begin{equation}
{\rm i}\hbar\,\frac{\partial \psi}{\partial t} = -\frac{\hbar^2}{2m}\Delta_g\psi+V\psi,
\end{equation}
where the Laplacian operator $\Delta_g$ is given by (sum in repeated indexes) 
\begin{equation}
\Delta_{g}\,\psi =\frac{1}{\sqrt{g}}\frac{\partial }{\partial q^i}\left(\sqrt{g}\,g^{ij}\frac{\partial\psi}{\partial q^j}\right), 
\end{equation}
$g$ being the determinant of the metric $g_{ij}$ in a coordinate system $(q^1,...,q^n)$ and $g^{ij}$ the coefficients of its inverse: $g^{ik}g_{kj}=\delta_j^i$.

In this work will be primarily concerned with the quantum dynamics on a surface or a curve in Euclidean space $\mathbb{R}^3$. In the following we introduce some basic notions concerning these objects that will be used throughout this article.

Here, $\Sigma$ always denotes a surface of class $C^k$, $k\geq3$. The coefficients of a metric $g_{ij}$ of $\Sigma$, also known as the first fundamental form, gives the line element  ${\rm d}s^2=g_{11}\,{\rm d}u^2+2g_{12}\,{\rm d}u\,{\rm d}v+g_{22}\,{\rm d}v^2$  and are obtained from a parametrization $\mathbf{x}(u,v)$ of $\Sigma$ as $g_{11}=\langle \mathbf{x}_u,\mathbf{x}_u\rangle$, $g_{12}=g_{21}=\langle \mathbf{x}_u,\mathbf{x}_v\rangle$, and $g_{22}=\langle \mathbf{x}_v,\mathbf{x}_v\rangle$, where $\mathbf{x}_u=\partial \mathbf{x}/\partial u$ and $\mathbf{x}_v=\partial \mathbf{x}/\partial v$ span the tangent plane $T_{p}\Sigma$ at $p=\mathbf{x}(u,v)$. The first fundamental form has to do with the intrinsic metric properties of the surface and any bending invariant can expressed as a function of theses coefficients only \cite{Manfredo,Struik}.

To obtain information on how the surface is (locally) embedded in the ambient space one introduces the second fundamental form $II=h_{11}\,{\rm d}u^2+2h_{12}\,{\rm d}u\,{\rm d}v+h_{22}\,{\rm d}^2v$. By introducing a normal vector field $\mathbf{N}(p)$ along $\Sigma$, e.g., $\mathbf{N}=(\mathbf{x}_u\times \mathbf{x}_v)/\,\Vert \mathbf{x}_u\times \mathbf{x}_v\Vert\perp T_p\,\Sigma$, the coefficients of the second fundamental form are given by $h_{11}=\langle \mathbf{x}_{uu},\mathbf{N}\rangle$, $h_{12}=\langle \mathbf{x}_{uv},\mathbf{N}\rangle$, and $h_{22}=\langle \mathbf{x}_{vv},\mathbf{N}\rangle$. From the first and second fundamental forms, the Gaussian and Mean Curvatures of $\Sigma$ are written as
\begin{equation}
K = \frac{h_{11}h_{22}-h_{12}^2}{g_{11}g_{22}-g_{12}^2}\,,\,H=\frac{1}{2}\frac{h_{11}g_{22}-2h_{12}g_{12}+h_{22}g_{11}}{g_{11}g_{22}-g_{12}^2},\label{eq::FormulaeKandH}
\end{equation}
respectively \cite{Manfredo,Struik}. It is possible to write the Gaussian curvature in terms of the coefficients of the first fundamental form only, which makes it a bending invariant. However, expressing it also using the second fundamental form can be more convenient in many cases. On the other hand, the same is not true for the Mean curvature, it is not a bending invariant and depends on the way the surface is embedded in the ambient space \cite{Manfredo,Struik}.

Besides their fundamental role in geometric considerations, the Gaussian and Mean curvatures appear in the quantum geometry-induced potential resulting from a confining procedure on a surface \cite{JensenKoppeAnnPhys,DaCostaPRA1981}:
\begin{equation}
V_{gip} = -\frac{\hbar^2}{2m}(H^2-K)\,.
\end{equation} 

By a regular curve we mean the image of a $C^k$ map $\alpha:I\to\mathbb{R}^3$, $k\geq3$, where $I\subseteq\mathbb{R}$ is an interval and $\Vert\alpha'(t)\Vert\not=0$ for all $t$. It is always possible to parametrize a curve $\alpha$ with a parameter $s$ in a way that $\Vert\dot{\alpha}(s)\Vert=1,$ known as the {\it arc-length} parametrization. We introduce the {\it unit tangent} vector field $\mathbf{t}(s)={\rm d}\alpha(s)/{\rm d}s$ along the curve and define the {\it curvature} function as $\kappa(s)=\Vert {\rm d}\mathbf{t}/{\rm d}s\Vert$ (it is a kind of scalar acceleration if we see $s$ as a time variable). Geometrically, the curvature function measures how much the curve deviates from being a straight line\footnote{Indeed, line segments are precisely the curves which satisfy $\kappa(s)\equiv0$.}. If we introduce the notion of an {\it osculating circle}, i.e. the best approximating circle for a curve in a given point, then it is possible to prove that $\kappa(s_0)=1/R,$ where $R$ is the radius of the osculating circle at $\alpha(s_0)$. We define the {\it principal normal} and the {\it binormal} vectors to be $\mathbf{n}=\mathbf{t}'/\Vert\mathbf{t}'\Vert$ and $\mathbf{b}=\mathbf{t}\times\mathbf{n}$, respectively. The vector fields $\{\mathbf{t},\textbf{n},\textbf{b}\}$ define the {\it Frenet trihedron} along the curve, and they satisfy the Frenet equations \cite{Struik} 
\begin{equation}
\frac{{\rm d}}{{\rm d}s}\left(
\begin{array}{c}
\textbf{t}\\
\textbf{n}\\
\textbf{b}\\
\end{array}
\right)=\left(
\begin{array}{ccc}
0 & \kappa & 0\\
-\kappa & 0 & \tau\\
0 & -\tau & 0\\
\end{array}
\right)\left(
\begin{array}{c}
\textbf{t}\\
\textbf{n}\\
\textbf{b}\\
\end{array}
\right),\label{eq::FranetFrameEqs}
\end{equation}
where $\tau$ is the the {\it torsion} of the curve $\alpha$, which measures how much the curve deviates from being planar\footnote{Indeed, a curve is planar if and only if $\tau\equiv0$.}. The curvature and torsion can be expressed as $\kappa(s) = \Vert\alpha''\Vert$ and $\tau(s)=\langle\alpha',\alpha''\times \alpha'''\,\rangle\Vert\alpha'\times \alpha''\Vert^{-2}$, respectively (if $\alpha$ is not parametrized by arc-length, the curvature is given as $\kappa=\Vert\alpha'\times\alpha''\Vert/\Vert\alpha'\Vert^{3}$ and the torsion is still given by the same expression). The Frenet equation uniquely determines a curve up to rigid motions, i.e., given two functions $\tilde{\kappa}(\tilde{s})>0$ and $\tilde{\tau}(\tilde{s})$, then integration of the Frenet equations above for an initial condition $\{\mathbf{t}_0,\textbf{n}_0,\textbf{b}_0\}$, formed by orthonormal vectors, determines a curve $\alpha$ with arc-length parameter $s=\tilde{s}$, curvature function $\kappa=\tilde{\kappa}$, and torsion $\tau=\tilde{\tau}$; the curve parametrization is given by $\alpha(s)=\alpha_0+\int_{s_0}^s\,\mathbf{t}(u){\rm d}u$.

Besides their fundamental role in geometric considerations for curves, the curvature appears in the quantum geometry-induced potential resulting from a 1D confining procedure\cite{JensenKoppeAnnPhys,DaCostaPRA1981}:
\begin{equation}
V_{gip} = -\frac{\hbar^2}{8m}\kappa^2\,.\label{eq::GIP1d}
\end{equation} 
Observe that the torsion does not enter the expression for the geometry-induced potential (this will be discussed in the following).

\subsection{Constrained Dynamics}

Let a mass $m$ in a space $M$ be confined to some $d$-dimensional region $N^d\subseteq M^{d+k}$ (the usual case being $M^{d+k}=\mathbb{R}^{d+k}$). Then, what are the ``correct" equations that govern the (constrained) dynamics on $N^d$? A first approach would be to use the intrinsic coordinates of $N^d$ and write the equations according to them\footnote{For example, the dynamics governed by a differential operator $L_M$ in $M$, such as the Laplacian $-\Delta_M$, is then described by the respective operator $L_N$ written on the $N^d$-coordinates.}. According to such an intrinsic scheme, the ambient space $M^{d+k}$ plays no relevant role at all. On the other hand, a different and more realistic approach would be to appeal to an explicit confining mechanism. In other words, one imposes that some kind of confining potential is responsible for the constraining, e.g., a strong force acting in the normal direction to $N$. Here the ambient space $M^{d+k}$ may play some role, since the confining potential ``sees'' the directions normal to  $N^d$, and then the constrained equations may depend on the way $N^d$ is embedded on $M^{d+k}$. Finally, one can also imagine a third different approach. Namely, one writes the equations in $M^{d+k}$ according to some coordinate system adapted to $N^d$, i.e., coordinates $(u^1,...,u^{d+k})$ such that $N^d=\{u\in M\,:\,u^{d+1}=u^{d+1}_0,...,u^{d+k}=u^{d+k}_0\}$ for some constants $u^{d+i}_0$, $i=1,...,k$, and then one takes the constrained dynamics on $N^d$ as the dynamics in $M$ after the last $k$ coordinates being fixed \footnote{We mention that, by the definition of a submanifold, it is always possible to find an adapted coordinate system in a certain
  neighborhood of a point of $N^d$; naturally, $\protect \mathaccentV
  {bar}016{u}=(u^1,...,u^d)\DOTSB \mapstochar \rightarrow
  (u^1,...,u^d,u^{d+1}_0,...,u_0^{d+k})\in M^{d+k}$,  for some constants $u^{d+i}_0$, $i=1,...,k$, is a (local)
  parametrization of $N^d$ into $M^{d+k}$.} : e.g., spheres in spherical coordinates. Generally, this approach is not equivalent to a confining potential one \cite{JensenKoppeAnnPhys,BernardAndLewYanVoonEurJPhys}. Indeed, since the equation $L_M(u)=0$, which describes the dynamics of the particle in $M^{d+k}$ according to a differential operator $L_M$, may involve derivatives with respect to $u^{d+1},...,u^{d+k}$, in general it does not follow that the solutions of $L_M(u\,;\{u^{d+i}=u^{d+i}_0\})$ are equivalent to the solutions of the respective operator $L_N(\bar{u})$ on $N$ written according to the adapted coordinate system.

In the classical mechanics picture, the approaches described above are shown to be equivalent, the choice between them being a matter of convenience. However, on the quantum mechanical counterpart, the dynamics must obey the uncertainty relations and, since any kind of confinement involves the fully knowledge of some degrees of freedom, it is not clear if  different approaches would lead to equivalent results for the constrained dynamics. We also mention that, by approaching the problem through a quantization procedure in the intrinsic coordinates of $N^d$, the resulting equations suffer from an ordering ambiguity \cite{DeWittRMP1957}. On the other hand, a confining potential approach does not suffer from such a problem: the confining potential approach gives a unique effective Hamiltonian to the confined dynamics \cite{DaCostaPRA1981}.

In the 1970s Jensen and Koppe\cite{JensenKoppeAnnPhys} showed how the many available approaches discussed above would lead to non-equivalent results through the illustrative example of a circle of radius $R$. More recently, Bernard and Lew Yan Voon \cite{BernardAndLewYanVoonEurJPhys} also discussed the non-equivalence for the case of spheroidal surfaces in $\mathbb{R}^3$, while Filgueiras {\it et al.} discussed the difference between intrinsic and confining potential approaches for conical surfaces \cite{BrasileirosJMP}.

In order to find the equations for the constrained dynamics in a surface $\Sigma\subset\mathbb{R}^3$, Jensen and Koppe \cite{JensenKoppeAnnPhys} devised an approach which consists in describing the confinement by starting from the dynamics in the region between two neighboring parallel surfaces and imposing homogeneous boundary conditions along them. So, taking the limit as the distance between the neighboring surfaces goes to zero, one obtains the equations that govern the constrained dynamics. Some years later, Da Costa devised an approach which consists in applying an explicit strong confining potential to restrict the motion of the particle to the desired surface (or curve) \cite{DaCostaPRA1981}. As expected, both formalisms coincide \cite{JensenKoppeAnnPhys,DaCostaPRA1981}; for surfaces one finds \cite{JensenKoppeAnnPhys,DaCostaPRA1981}    
\begin{equation}
{\rm i}\hbar\frac{\partial \psi}{\partial t} = -\frac{\hbar^2}{2m}\left[\Delta_{\Sigma}+(H^2-K)\right]\psi\,;\label{eq:SchrodingerEqExtrinsic}
\end{equation}
while for curves one has \cite{DaCostaPRA1981}
\begin{equation}
{\rm i}\hbar\frac{\partial \psi}{\partial t} = -\frac{\hbar^2}{2m}\left(\Delta_{\alpha}+\frac{\kappa^2}{4}\right)\psi,\label{eq:SchrodingerEqExtrinsicCurves}
\end{equation}
where $\Delta_{\alpha}={\rm d}^2/{\rm d}s^2$ is the Laplacian on the curve in terms of its arc-length parameter (see subsection 3.1 below). The above equations show that in general the intrinsic and confining potential approaches do not lead to the same constrained dynamics. In the former, the dynamics is governed by the Laplacian operator only, while in the latter the Laplacian is coupled to a scalar geometry-induced potential. So, in order to do a more realistic study, where the global geometry should be taken into account, an extrinsic scheme would be more appropriate. Additionally, the equations will be exactly the same only for (regions) of the plane or spheres, since these are the only surfaces where $H^2-K=0$, while the equality for curves occurs uniquely for line segments, since it is demanded $\kappa^2\equiv0$.

Finally, it is worth to mention that these results for the constrained dynamics are based on the assumption that the confining potential $V_{\mathrm{conf}}$ is uniform, i.e., its equipotentials only depend on the distance from the constraint region $N\subseteq M$: $V_{\mathrm{conf}}(q)=V_{\mathrm{conf}}(\mbox{dist}(q,N))$. The confinement is then put forward through a limiting procedure, i.e., one considers a sequence of potentials $\{V_{\lambda}\}_{\lambda}$ that approximates the confining one $V_{\mathrm{conf}}$ for $\lambda\to\infty$ \cite{DaCostaPRA1981}: 
\begin{equation}
V_{\mathrm{conf}}(q)=\lim_{\lambda\to\infty} V_{\lambda}(q) = \left\{
\begin{array}{ccc}
0 & , & q\in N\\
\infty & , & q\not\in N\\
\end{array}
\right.,
\end{equation}
which allows for the decoupling between the tangential and normal degrees of freedom in the limit $\lambda\to\infty$. So, one separates the Hamiltonian into a term that governs the low energy motion in the tangent direction, which is the effective Hamiltonian along the constraint region, and a high energy motion in the normal direction. However, in some context this hypothesis is no longer realistic and one can not suppose that the equipotentials are equidistant. As a consequence, the tangential and normal degrees of freedom are coupled \cite{WachsmuthPRA2010,SchmelcherPRA2014}. In what follows we will not consider such a possibility.

Let us finish this section by making some remarks concerning the role played by the torsion for curves. Interestingly, the torsion of a curve does not appear in the GIP \cite{DaCostaPRA1981}. Nonetheless, Takagi and Tanzawa put forward an investigation for a particle confined to a thin tube, which is twisted and curved to form a closed loop \cite{TwistedRingProgTheorPhys}, and described the effect of both curvature and torsion of the loop up to second order. They then observed that the torsion may give rise to a geometry-induced Aharonov-Bohm effect. On the other hand, in the study of the Schr\"odinger-Pauli equation for a spin-orbit coupled electron constrained to a space curve \cite{OrtixPRB2015}, it was found that the torsion of the curve generates an additional quantum geometry-induced potential, adding to the known curvature-induced one. In short, besides making the integration of the Frenet equations more difficult (see section below), these studies suggest that by considering other effects, in addition to the constraining for the Schr\"odinger equation, the torsion naturally appears in the discussion. Moreover, by noticing that the torsion has to do with the derivative of the binormal vector $\mathbf{b}$, which can be expressed as $\mathbf{b}=(\alpha'\times\alpha'')/\Vert\alpha'\times\alpha''\Vert$ \cite{Manfredo,Struik}, one would say that the torsion is somehow related to an angular momentum. So, it seems natural to expect that the torsion appears in those contexts where the angular momentum plays a role.

\section{Curves and Surfaces with Prescribed Geometry-induced Potential}

Exploiting the effects of an extra contribution to the Hamiltonian due to a confining potential approach is essential and in this respect an important problem is that of a prescribed geometry-induced potential, i.e., the inverse problem of finding a curved region with a potential given {\it a priori}. The solution of this problem offers the possibility of engineering surfaces and curves with a quantum behavior prescribed {\it a priori} through their geometry-induced potential and has been already investigated for curves \cite{delCampoSciRep2014} and a class of revolution surfaces \cite{AtanasovPLA2007}.

For the confinement on a curve $\alpha:I\to\mathbb{R}^3$ the problem reduces to that of finding a curve with a prescribed curvature function (\ref{eq::GIP1d}). The curve is then obtained after integration of the Frenet equations (\ref{eq::FranetFrameEqs}) as $\alpha(s)=\alpha_0+\int_{s_0}^s\,\mathbf{t}(u)\,du.$ It is worth to mention that in the case of planar curves, i.e., $\tau\equiv0$, the parametrization of the solution curve for the Frenet equations are \cite{Struik}
\begin{equation}
\left\{
\begin{array}{c}
x(s)=z_{1}\,C(s)-z_{2}\,S(s)+x_0\\[6pt]
y(s)=z_{1}\,S(s)+z_{2}\,C(s)+y_0\\
\end{array}
\right.\,,
\end{equation}
where $x_0,\,y_0,$ and $z_{i}$ are constants to be specified by the initial conditions and
\begin{equation}
\left\{
\begin{array}{c}
S(s)=+\displaystyle\int_{s_0}^s\cos\Big(\int_{s_0}^v\kappa(u)\,du\Big)dv\\[8pt]
C(s)=-\displaystyle\int_{s_0}^s\sin\Big(\int_{s_0}^v\kappa(u)\,du\Big)dv\\
\end{array}
\right.\,.\label{eq::SenoCosFrenetFrame}
\end{equation}
Recently, del Campo {\it et al.} \cite{delCampoSciRep2014} exploited such an explicit solution in order to find pair of curves with the same scattering properties. Finally, for $\tau\not=0$, it is possible to find the general solution for the Frenet equations by writing them in term of a complex Riccati equation \cite{Struik}.
\newline

For surfaces, the situation is more complex. Indeed, the prescribed GIP problem generally demands the solution of a 2nd order non-linear PDE. For example, assuming the surface to be the graph of a smooth function $Z(x,y)$, i.e., the parametrization is given by $\mathbf{r}(x,y) = (x,y,Z(x,y))$, the Gaussian and Mean curvatures are written as
\begin{equation}
K(x,y) = \displaystyle\frac{Z_{xx}Z_{yy}-Z_{xy}^2}{(1+Z_{x}^2+Z_{y}^2)^2}=\frac{\det(\mbox{Hess}\,Z)}{(1+\Vert\nabla Z\Vert^2)^2}
\end{equation}
and
\begin{equation}
H(x,y) = \displaystyle\frac{Z_{xx}(1+Z_{y}^2)-2Z_{xy}Z_{x}Z_{y}+Z_{yy}(1+Z_{x}^2)}{2(1+Z_{x}^2+Z_{y}^2)^{3/2}}=\frac{1}{2}\nabla\cdot\left(\frac{\nabla Z}{\sqrt{1+\Vert\nabla Z\Vert^2}\,}\right),
\end{equation}
respectively. The equation for $K$ is a nonlinear elliptic PDE of Hessian type (also referred as Monge-Amp\`ere equation) \cite{Gutierrez2001}, while the equation for $H$ is a nonlinear elliptic PDE of divergent type \cite{GilbargTrudinger1977}. 

A general study of the PDE associated with the prescribed GIP $H^2-K$ is not a trivial task. In addition it can encode in its generality useless examples. In this respect, the study of particular classes can turn to be more useful and insightful than a general analysis. So, instead of studying the prescribed potential problem in general, which would lead us to the realm of non-linear analysis \cite{Gutierrez2001,GilbargTrudinger1977}, here we restrict ourselves to the simpler, but still important and difficult, context of invariant surfaces (continuous symmetries). To be more precise, we assume the surfaces to be invariant by a 1-parameter group of isometries of $\mathbb{R}^3$ \cite{DoCarmoTohoku1982,MedeirosRMU1991}. This allows us to avoid the study of a non-linear PDE, since the symmetry turns the equation into an (non-linear) ODE along the so called generating curve. 

\subsection{Surfaces invariant by a 1-parameter subgroup of isometries}

Basically there exist three types of surfaces invariant by a 1-parameter subgroup of isometries of $\mathbb{R}^3$, namely (i) cylindrical surfaces (translation symmetry), (ii) surfaces of revolution (rotational symmetry), and (iii) helicoidal surfaces (screw rotation symmetry, i.e., a combination of a translation and a rotation). Due to their appealing symmetry, these surfaces are commonly encountered in applications and theoretical studies in the context of a constrained dynamics: e.g., cylindrical surfaces to model rolled-up nanotubes \cite{OrtixPRB2010} and $\pi$ electron energies of aromatic molecules \cite{BastosEtAlPreprint,MiliordosPRA2010}; surfaces of revolution as tractable examples to test the validity and potentialities of an extrinsic confinement approach \cite{EncinosaPRA,GravesenJMathPhys}; and helicoidal surfaces to study geometry-induced charge separation \cite{AtanasovPRB2009} and the relation to the concept of chirality \cite{AtanasovPRB2015}, just to name a few.
\newline

A function $T:\mathbb{R}^3\to\mathbb{R}^3$ is an isometry of $\mathbb{R}^3$ if it satisfies for all $q,\tilde{q}\in\mathbb{R}^3$ the relation $\langle T(q),T(\tilde{q})\rangle=\langle q,\tilde{q}\rangle$. These functions form the so called group of rigid motions of $\mathbb{R}^3$, which are composed by translations $T_a(q)=q+a$ and rotations $R\in O(3)$ (or $SO(3)$ if one imposes that $T$ preserves orientation). By a one-parameter subgroup of isometries we mean an action of the additive group $(\mathbb{R},+)$ on the symmetry group (rigid motions) of $\mathbb{R}^3$. In other words, a 1-parameter subgroup of isometries is a smooth map $\gamma:\mathbb{R}\times\mathbb{R}^3\to\mathbb{R}^3$ such that
\begin{enumerate}[(a)]
\item For all $t\in\mathbb{R}$ the map $q\mapsto \gamma(t,q)$, denoted by $\gamma_t$, is a rigid motion;
\item For all $t,s\in\mathbb{R}$, $\gamma_t\circ\gamma_s=\gamma_{t+s}$ and $\gamma_0=Id$ is the identity map.
\end{enumerate}
Up to a change of variables, every 1-parameter subgroup can be written as \cite{MedeirosRMU1991}
\begin{equation}
\gamma(t,q)=(q^1\cos t+q^2\sin t,-q^1\sin t+q^2\cos t,q^3+ht),
\end{equation}
or as
\begin{equation}
\gamma(t,q)=(q^1,q^2,q^3+ht),
\end{equation}
where $h\in\mathbb{R}$ is a constant, equal to zero for rotational symmetry in the former or equal to zero for the identity map in the latter.
\newline
\newline
{\it Remark:} When discussing the constrained dynamics on a helicoidal surface it will prove useful to adopt a different notation. More precisely, we will assume the 1-parameter subgroup of isometries to be $$\gamma(t,q)=(q^1\cos (\omega t)+q^2\sin(\omega t),-q^1\sin(\omega t)+q^2\cos(\omega t),q^3+t),$$ where $\omega$ is a constant.
\newline

A surface $\Sigma\subseteq \mathbb{R}^3$ invariant by a 1-parameter subgroup of isometries of $\mathbb{R}^3$ is characterized by 
\begin{equation}
\forall\,t\in\mathbb{R},\,\Sigma = \gamma_t(\Sigma)\,.
\end{equation}
Intuitively, we can approximate an invariant surface by successive applications, to a given curve $\alpha(s)$, of a certain kind of rigid motion: $$\Sigma\cong\{\gamma_{t_0}(\alpha(s)),\gamma_{t_0+\Delta t}(\alpha(s)),\cdots,\gamma_{t_0+n\Delta t}(\alpha(s))\}.$$ So, in the limit $\Delta t\to 0$, we generate the surface by continuously moving the curve $\alpha$ by the action of a 1-parameter subgroup $\gamma_t$. We call such a curve the {\it generating curve}, which can be assumed to be planar.

It follows that the values of the Gaussian and Mean curvature only depend on the values assumed along the generating curve. As a corollary of the invariance of $K$ and $H$, the prescribed GIP problem demands the solution of an ODE instead of a PDE. In the next section we present a study of this problem for each type of invariant surface. 

\section{Cylindrical surfaces with prescribed geometry-induced potential}

Now we focus on the simplest instance of surfaces invariant by a 1-parameter subgroup of isometries, namely, surfaces with translation symmetry. A cylinder is the standard example, it is just the surface obtained by translating a circle. More generally, a cylindrical surface is obtained by taking a generating curve (the cross section) which can be any planar curve $\alpha(s):I\to\mathbb{R}^2\subset\mathbb{R}^3$ (for a study of cylindrical surfaces with a varying cross section see \cite{BastosEtAlPreprint}). We then translate this curve in the direction of a unit vector ${\bf a}=(a_1,a_2,a_3)$, where we assume $a_3\not=0$ in order to have a regular surface, i.e., ${\bf a}$ is out of the $xy$ plane\footnote{We could have assumed $\mathbf{a}$ to be $(0,0,1)$, but we decided to work with an arbitrary vector in order to include inclined cylinders in our discussion.}. By denoting $\alpha(s)=(x(s),y(s),0)$, where $s$ is an arc-length parameter, we have the following parametrization for a cylindrical surface
\begin{equation}
X(s,t) = \alpha(s)+t\,{\bf a}.\label{eq::parCylindricalSurface}
\end{equation}
Observe that the generating curve does not need to be closed.

The coefficients of the first and second fundamental forms are given by
\begin{equation}
g_{11}(s,t)=1,\,g_{12}(s,t)=\cos\theta,\,g_{22}(s,t)=1,
\end{equation}
and
\begin{equation}
h_{11}(s,t)=h_{12}(s,t)=0,h_{22}(s,t)=\langle\alpha'\times\alpha'',\mathbf{a}\rangle,
\end{equation}
respectively; where we have adopted the unit normal ${\bf n}=\mathbf{a}\times\alpha'$ and $\theta=\cos^{-1}\langle \alpha',\mathbf{a}\rangle$  is the (constant) angle between $\mathbf{a}$ and $\alpha'$. Now we can compute the Gaussian and Mean curvatures of a cylindrical surface as
\begin{equation}
K\equiv0\,\mbox{ and }\,H=\frac{a_3[x'(s)\,y''(s)-x''(s)\,y'(s)]}{2\,\sin^2\theta}\,,
\end{equation}
respectively. Notice, as expected, that due to the translation symmetry the Gaussian and Mean curvatures are functions of $s$ only. On the other hand, since $K\equiv0$, the problem of a prescribed GIP $H^2-K$ is equivalent to the problem of finding cylindrical surfaces with prescribed Mean curvature. Then, given a function $H(s)$, one must solve the following system of 2nd order nonlinear ODEs
\begin{equation}
\left\{
\begin{array}{c}
x'\,y''-x''\,y'=\frac{2\,{\sin^2\theta}}{a_3}\,H(s)\\[5pt]
(x')^2+(y')^2=1\\
\end{array}
\right.\,,\label{eq::ODEPresc_H_ForCylindricalSur}
\end{equation}
where the second equation comes from the parametrization by arc-length. 

For a planar curve $\alpha(s)=(x(s),y(s))$, we can write the curvature function as \cite{Manfredo,Struik}
\begin{equation}
\kappa = \frac{x'\,y''-x''\,y'}{[(x')^2+(y')^2]^{3/2}}.
\end{equation}
Then, we have the following result 
\begin{prop} 
The Mean curvature $H(s)$ of a cylindrical surface and the curvature function $\kappa(s)$ of its generating curve (cross section) are related according to
\begin{equation}
\kappa(s) = \frac{2\,{\sin^2\theta}}{a_3}H(s)\,,\label{eq::RelationCurvCurveAndHcylindSurf}
\end{equation}
where $\theta$ is the (constant) angle between the direction of translation $\mathbf{a}=(a_1,a_2,a_3)$ and the plane which contains the generating curve. Moreover, it follows that Eq. (\ref{eq::FranetFrameEqs}) solves the problem of prescribed Mean curvature, i.e., there is an equivalence between finding curves with prescribed curvature and finding cylindrical surfaces with prescribed Mean curvature.
\end{prop}



\section{Surfaces of revolution with prescribed geometry-induced potential}


A first attempt to solve the prescribed GIP problem for surfaces of revolution was devised by Atanasov and Dandoloff \cite{AtanasovPLA2007}. They considered surfaces of revolution whose generating curve, to be rotated around the $z$ axis, is a graph on the $xz$ plane. They also investigated the existence of bound states and surfaces in the form of circular strips around the symmetry axis.

In the following, we consider surfaces of revolution without imposing any restriction on the generating curve. We show that the equation for the prescribed GIP can be rewritten as a first order complex equation. Further, we specialize to surfaces whose generating curve is a graph on the $xz$ plane that can be rotated around either the $x$ or the $z$ axis. 

Given a curve $\alpha(s)=(x(s),0,z(s))$ on the $xz$ plane ($s$ being its arc-length parameter), the surface of revolution obtained by rotating $\alpha$ around the $z$ axis is parametrized by
\begin{equation}
x(s,\phi)=(x(s)\cos\phi,x(s)\sin\phi,z(s)),
\end{equation}
where we must assume $x(s)>0$ for all $s$.

The coefficients of the first and second fundamental forms are given by
\begin{equation}
g_{11} = 1,\,\,g_{12}=0,\,\,g_{22}=x^2(s),
\end{equation}
and
\begin{equation}
h_{11} = x'(s)z''(s)-x''(s)z'(s),\,\,h_{12}=0,\,\,h_{22}=x(s)z'(s),
\end{equation}
respectively. From these expressions we find
\begin{equation}
U=\sqrt{H^2-K} = \frac{x(x'z''-x''z')-z'}{2x}.\label{eqPrescribedGIPFullRevolSurf}
\end{equation}
Observe the similarity of this expression with that of the Mean curvature:
\begin{equation}
H = \frac{x(x'z''-x''z')+z'}{2x}.
\end{equation}
Indeed, they are the same except for the exchange of the sign in front of $z'$. This similarity will be exploited in the following.

The equation of prescribed $U$, or $H$, is a 2nd order non-linear ODE\footnote{In fact, since we are assuming $x'^2+z'^2=1$, the prescribed curvature problem is given by a system of 2nd order non-linear ODE's.}. In the 80's, Kenmotsu solved the prescribed Mean curvature equation by transforming it in a 1st order complex linear ODE \cite{KenmotsuTohokuMathJ}: $Z'-2\,{\rm i}\,H\,Z+1=0$. This technique can be applied to our problem, i.e., we can write the equation for $U$ as a 1st order complex ODE, which in our case is non-linear: $Z'-2\,{\rm i}\,U\,Z+\vert Z\vert^2=0$.

Multiplying Eq. (\ref{eqPrescribedGIPFullRevolSurf}) by $x'$, and using $x'^2+z'^2=1$ and its derivative, we find
\begin{equation}
0=2xx'U+x'z'-xz''=2\frac{x'}{x}U-\left(\frac{z'}{x}\right)'\,.
\end{equation}
On the other hand, multiplying Eq. (\ref{eqPrescribedGIPFullRevolSurf}) by $z'$, and using $x'\,^2+z'\,^2=1$ and its derivative, we have
\begin{equation}
0=2xz'U+xx''-x'\,^2+1=2\frac{z'}{x}U+\left(\frac{x'}{x}\right)'+\frac{1}{x^2}\,.
\end{equation}
Finally, defining $Z(s)=x^{-1}(s)[x'(s)+{\rm i}\,z'(s)]$, we can glue the above equations together and write
\begin{equation}
Z'(s)-2\,{\rm i}\,U(s)\,Z(s)+\vert Z(s)\vert^2=0\,.
\end{equation}

In the next subsections we will study some particular classes of revolution surfaces where the equation for the prescribed potential can be effectively solved. 

\subsection{Surfaces whose generating curve is a graph rotated around a vertical axis}
 
In the end of the 1990s, Baikoussis and Koufogiorgos \cite{BaikoussisJGeom} studied the problem of finding helicoidal surfaces with prescribed Mean or Gaussian curvatures. They assumed a parametrization
given by
\begin{equation}
x(\rho,\phi) = (\rho\cos \phi,\rho\sin \phi,\lambda(\rho)+h\,\phi),\,\rho>0,\label{eq::ParamHelicoidalSur}
\end{equation}
where $h$ is a constant and $\lambda(\rho)$ a smooth function, which represents the generating curve $(\rho,0,\lambda(\rho))$. As natural, $\phi$ stands for the rotation angle around the $Oz$ axis, the screw axis, and $\rho$ for the distance from it. 

If $h=0$, the helicoidal surface is just a surface of revolution, while if $\lambda\equiv0$ and $h\not=0$ one has the usual helicoid surface. In addition, since the generating curve $\lambda$ is supposed to be a graph, cylinders are not covered by (\ref{eq::ParamHelicoidalSur}) (such an example will be covered in the following subsection by allowing a rotation around the $x$ axis).  

The problem of prescribed Mean or Gaussian curvatures is then solved by writing the curvatures of the given surface in terms of the parameters $h$ and $\lambda(\rho)$. This leads to an ODE that, if properly manipulated, can be written as 
\begin{equation}
\frac{\rho}{2}A'(\rho)+A(\rho)=H_0(\rho) \mbox{ and }\frac{1}{2\rho}(B^2(\rho))'=K_0(\rho),\label{eq::MeanAndGaussODE}
\end{equation}
where
\begin{equation}
A = \frac{\lambda'}{\sqrt{\rho^2(1+\lambda'\,^2)+h^2}}\,;\,B^2=\frac{\rho^2\lambda'\,^2+h^2}{\rho^2(1+\lambda'\,^2)+h^2}\,.\label{eq::AandBfromMeanGaussEDO}
\end{equation}

We now apply these ideas to surfaces of revolution by imposing  $h=0$. It follows that $B^2=\rho^2A^2$, which gives us the following ODE in terms of $U$ ($=\sqrt{H^2-K}$)
\begin{equation}
\frac{\rho^2}{4}(A')^2=U^2\Rightarrow A(\rho)=\pm\left(2\int U(\rho)\,\frac{{\rm d}\rho}{\rho}+a_1\right)\,,\label{eq::willmoreEDO}
\end{equation}
where $a_1$ is a constant of integration. Using this in Eq. (\ref{eq::AandBfromMeanGaussEDO}) under the condition $h=0$, one obtains an ODE for the generating curve $\lambda(\rho)$:
\begin{equation}
\lambda'\,^2 = A^2\,\rho^2\,(1+\lambda'\,^2)\Rightarrow [1-\rho^2A^2]\,\lambda'\,^2=\rho^2\,A^2\geq0.
\end{equation}
By continuity, if $1-\rho_0^2A(\rho_0)>0$ at some $\rho_0\in\mathbb{R}-\{0\}$, then $1-\rho^2A^2(\rho)>0$ on a neighborhood of $\rho_0$. So, one gets the general solution in the neighborhood of $\rho_0$
\begin{equation}
\lambda(\rho) = \pm\int\frac{\rho A(\rho)}{\sqrt{1-\rho^2A^2(\rho)}}\,{\rm d}\rho+a_2,\label{eq::CurvaSolucaoPotPrescrito}
\end{equation}
where $A(\rho)$ is given by Eq (\ref{eq::willmoreEDO}) and $a_2$ is another constant of integration.

In short, given a smooth function $U(\rho)$, we can define a 2-parameter family of curves
\begin{equation}
\gamma(\rho;U(\rho),a_1,a_2) = \pm\int\displaystyle\frac{\rho\,\Big(2\int\,U\,\frac{{\rm d}\rho}{\rho}+a_1\Big)}{[1-\rho^2(2\int\,U\,\frac{{\rm d}\rho}{\rho}+a_1)^2]^{1/2}}\,{\rm d}\rho+a_2\,.\label{eq::familyCurvesLambda}
\end{equation}
which furnishes a 2-parameter family of surfaces of revolution with a GIP $\sqrt{H^2(\rho)-K(\rho)}=U(\rho)$ by applying a rotation around the $z$-axis.
\newline
\newline
{\it Example 5.1: (vanishing geometry-induced potential)} For $U\equiv0$, Eq. (\ref{eq::willmoreEDO}) gives $A(\rho)=a_1$ constant and, from Eq. (\ref{eq::familyCurvesLambda}), one has 
\begin{equation}
\lambda(\rho) =\left\{
\begin{array}{ccc}
 \pm\sqrt{a_1^{-2}-\rho^2}+a_2 & , & a_1\not=0\\
 a_2 & , & a_1 = 0\\
\end{array}
\right..
\end{equation}
Then, for $a_1\not=0$, one has a sphere of radius $R=1/a_1$, and if $a_1=0$ one has a region of a plane. By a well known result, the only surfaces satisfying $H^2-K\equiv0$ are (pieces of a) sphere or plane (see Ref. \cite{Manfredo}, p. 147). In this way we recovered the two cases of surfaces where $H^2-K\equiv0$.
\newline
\newline
{\it Example 5.2: (constant geometry-induced potential)} Remember that for a cylinder of radius $R$, the geometry-induced potential is $U\equiv(2R)^{-1}$. However, a cylinder can not be obtained from the parametrization in Eq. (\ref{eq::ParamHelicoidalSur}); for a cylinder $x(\rho,\phi)=(R\,\cos \phi,R\,\sin \phi,\rho)$. Now we show that there are other examples of surfaces of revolution, which are not a cylinder, with $U\equiv U_0\not=0$ constant. The importance of such examples lies in the fact that surfaces with a constant GIP have the same set of eigenfunctions of the problem without the GIP\footnote{Indeed, two Hamiltonians $\hat{H}_i=-\hbar^2/2m\,\Delta_g+\mathbb{G}_i$ differ by a constant, i.e., $\mathbb{G}_1-\mathbb{G}_2\equiv$ constant, if and only if they have the same set of eigenfunctions when subjected to the same boundary conditions. In this case, if $E^{(1)}_n$ and $E^{(2)}_n$ denote the respective eigenvalues for the same eigenfunction $\psi_n$, we have $E^{(1)}_n-E^{(2)}_n = \mathbb{G}_1-\mathbb{G}_2$ (notice that the gap between the eigenvalues satisfies $E_{n+k}^{(2)}-E_{n}^{(2)}=E_{n+k}^{(1)}-E_{n}^{(1)}$).}. 

Indeed, assuming $U(\rho)=U_0$ constant, Eq. (\ref{eq::familyCurvesLambda}) gives 
\begin{equation}
\lambda(\rho)=\pm\int_{\rho_0}^{\rho}\frac{x\Big(2\,U_0\ln\left(\frac{x}{\rho_0}\right)+a_1\Big)}{[1-x^2(2\,U_0\ln\left(\frac{x}{\rho_0}\right)+a_1)^2\,]^{1/2}}\,{\rm d}x+a_2\,. 
\end{equation} 
The rotation of this curve around the $z$ axis generates a non-cylindrical surface with constant GIP $U_0$.

\subsection{Surfaces whose generating curve is a graph rotated around a horizontal axis}
 
Now we focus on another class of surfaces of revolution. In the previous analysis, the curve on the $xz$ plane to be rotated around the $z$ axis was supposed to be a graph, i.e., of the form $z=z(x)$. In this way, the surfaces obtained do not include cylinders and, more generally, do not include the surface of deformed nanotubes \cite{FernandosEfranceses} also. To include such examples, we can enlarge our class of surfaces by allowing a rotation of a curve $z=z(x)$ around the $x$ axis. We can parametrize these surfaces according to
\begin{equation}
x(q,\phi) = (q,\rho(q)\sin\phi,\rho(q)\cos\phi),
\end{equation}
where $\rho(q)>0$ is a function which represents the distance to the rotation axis and defines the generating curve $(q,0,\rho(q))$ in the $xz$ plane to be rotated around the $x$ axis. As usual, $\phi$ is the angle of rotation. 

The geometry-induced potential of such surfaces can be written as \cite{FernandosEfranceses}
\begin{equation}
V_{gip}=-\frac{\hbar^2}{2m}\frac{[1+\rho'(q)^2+\rho(q)\rho''(q)]^2}{4\rho(q)^2[1+\rho'(q)^2]^{3}}\,,\label{eq::PotGeomTuboEnrugado}
\end{equation}
which furnishes for $U=\sqrt{H^2-K}$ the expression
\begin{equation}
\pm U=\frac{1+\rho'(q)^2+\rho(q)\rho''(q)}{2\rho(q)[1+\rho'(q)^2]^{3/2}}=-\frac{\rho}{2\rho'}\frac{{\rm d}A}{{\rm d}q},
\end{equation}
where
\begin{equation}
A=\frac{1}{\rho(q)[1+\rho'(q)^2]^{1/2}}\,.
\end{equation}
Then, we have the following differential equation for $A$
\begin{equation}
\rho\frac{{\rm d}A}{{\rm d}q}+2(\pm U)\frac{{\rm d}\rho}{{\rm d}q}=\left[\rho\frac{{\rm d}A}{{\rm d}\rho}+2(\pm U)\right]\frac{{\rm d}\rho}{{\rm d}q}=0.
\end{equation}
If $\rho'\equiv0$, then $\rho=$ constant and we have a cylinder. Otherwise, we find the following ODE in terms of $\rho$
\begin{equation}
\rho\frac{{\rm d}A}{{\rm d}\rho}+2(\pm U)=0\Rightarrow A(\rho)=\pm\left(2\int\frac{{\rm d}\rho}{\rho}\,U(\rho)+a_1\right)\,,
\end{equation}
where $a_1$ is a constant of integration. Notice that this last equation is identical to Eq. (\ref{eq::willmoreEDO}), with the difference that here $\rho=\rho(q)$ is the function that we are trying to find.

Now, by using the definition of $A$, we find
\begin{equation}
\frac{d\rho}{dq}=\pm\sqrt{\frac{1-\rho^2A^2}{\rho^2A^2}}\Rightarrow q(\rho)=\pm\int \frac{\rho A}{\sqrt{1-\rho^2A^2}}\,{\rm d}\rho+q_0\,.
\end{equation}
This equation is identical to Eq. (\ref{eq::CurvaSolucaoPotPrescrito}), but instead of obtaining the function which gives the generating curve, we obtained its inverse. This result reveals a certain duality between the surface of revolution obtained by rotating a curve $z=z(x)$ around the $x$ or the $z$ axes. In other words, 

\begin{prop}
Let $U$ be a smooth function of one variable, then each curve of the 2-parameter family given in (\ref{eq::CurvaSolucaoPotPrescrito}) generates a surface of revolution whose geometry-induced potential is $\sqrt{H^2-K}=U$ when rotated around the $x$ or the $z$ axis.
\end{prop}

\section{Helicoidal surfaces}

The definition of chirality comes from the fact that some objects can not be transformed into their mirror image under applications of rigid motions. This idea is present in many scientific areas and is of fundamental importance \cite{CahnAngewChem1966}. It appears in nature, such as tendrils and gastropod shells, and more fundamentally in the structure of DNA molecules. The study of chiral molecules is an important branch of stereochemistry with many applications in inorganic, organic, and physical chemistry, and also with several implications for the pharmaceutical industry. The concept of chirality is also present in particle physics and condensed matter \cite{KondepudiSciAmer1990}. In particular, this concept has proved to be useful in understanding some recent experimental results related to electronic, mechanical, and optical properties of nanotubes \cite{Yakobson2014}.

Recently, a link between chirality and the constrained particle dynamics was observed in the study of a particle on a helicoid \cite{AtanasovPRB2009, AtanasovPRB2015}. A helicoid is a particular instance of a helicoidal surface. These surfaces form the natural candidates to investigate a link with the concept of chirality. Indeed, given a curve $\alpha(\rho)=(\rho,0,\lambda(\rho))$ on the $xz$ plane, we can obtain enantiomorphic surfaces by screw-rotating $\alpha$ around the $z$ axis clock and counterclockwisely:
\begin{equation}
(\rho\cos(\omega\phi),\rho\sin(\omega\phi),\lambda(\rho)+\phi)\leftrightarrow(\rho\cos(\omega\phi),-\rho\sin(\omega\phi),\lambda(\rho)+\phi)\,.
\end{equation}
Observe that the sign of the constant $\omega$ can be used in order to control the chirality of the respective surface.

In the following, we study the geometric properties of helicoidal surfaces and comment on the existence of the so-called natural parameters, which allows for a better understanding and unified approach to such surfaces. The study of the respective Schr\"odinger equation under the influence of the GIP in such a coordinate system, along with some comparisons with known results for the dynamics on a helicoid, are present in the next section.  

\subsection{Parametrization by natural parameters}

Helicoidal surfaces are invariant by a rotation in combination with a translation (screw-rotation), the standard example being a helicoid, whose generating curve is just a line segment $(\rho,0,0)$:
\begin{equation}
X_{helic}(\rho,\phi) = (\rho\cos(\omega\phi),\rho\sin(\omega\phi),\phi),\label{eq:ParHelicoid}
\end{equation}
where $\omega$ is a constant. If $L$ is the height of the helicoid, then we can write $\omega=2\pi n/L$, where $n$ is the number of twists around the screw-rotation axis. Moreover, the sign of $\omega$ governs the distinct chiralities states exhibited by helicoidal surfaces and has some consequences for the dynamics \cite{AtanasovPRB2009,AtanasovPRB2015}.
\newline
\newline
{\it Remark:} In the previous Section we have already encountered helicoidal surfaces, Eq. (\ref{eq::ParamHelicoidalSur}), but here we adopt a different notation in order to ease comparisons with known results for the dynamics on a helicoid. As a consequence, surfaces of revolution are not allowed, since a translation in the direction of the screw axis is always present. However, surfaces of revolution can be formally obtained by changing $\omega\phi\mapsto \phi$ and then taking $\omega\to\infty$.
\newline

For the helicoid, the coordinate system $(\rho,\phi)$ allows for a simple interpretation: $\phi$ represents the rotation angle (observe that the translation in the direction of the screw axis is proportional to the angular rotation), while the $\rho$-constant curves are helices; $\rho$ is the distance from the screw axis. On the other hand, for a general helicoidal surface, the translation along the screw axis has an extra contribution, which depends on the height of the generating curve $\alpha(\rho)=(\rho,0,\lambda(\rho)),$ $\rho>0$. Then, we have
\begin{equation}
X(\rho,\phi) = (\rho\cos(\omega\phi),\rho\sin(\omega\phi),\lambda(\rho)+\phi).\label{eq:ParHelicSurf}
\end{equation}

In the above parametrization of a helicoidal surface the coordinate system $(\rho,\phi)$ does not have the same interpretation as happens for a helicoid. Indeed, in order to achieve that one could use a coordinate system composed of \textit{natural parameters}. More precisely, we say that a helicoidal surface $\Sigma$ is parametrized by {\it natural parameters} $(\xi,\chi)$ if:
\begin{enumerate}[(i)]
\item $\xi$-curves ($\chi$ constant) are parametrized by the arc-length parameter; and
\item $\chi$-curves ($\xi$ constant) are helices orthogonal to the $\xi$-curves. 
\end{enumerate}
In other words, since $\xi$ is the arc-parameter of a $\chi$-curve, the parameter $\xi$ represents a distance from the screw axis, while $\chi$ denotes the parameter along the orbits of the screw rotation symmetry, i.e., helices. This is precisely what happens for a helicoid, where $\xi_{helic}=\rho$ and $\chi_{helic}=\phi$ (here $\lambda_{helic}\equiv0$).

A useful consequence of using natural parameters is that the metric can be written in a simpler form:
\begin{equation}
{\rm d}s^2 = {\rm d}\xi^2+\mathcal{U}^2(\xi)\,{\rm d}\chi^2,
\end{equation}
for some function $\mathcal{U}$.

 It is possible to show that every helicoidal surface admits a reparametrization by natural parameters \cite{DoCarmoTohoku1982}. Indeed, from the line element
\begin{eqnarray}
{\rm d}s^2 & = & (1+\lambda'\,^2){\rm d}\rho^2+2\lambda'{\rm d}\rho\,{\rm d}\phi+(1+\omega^2\rho^2){\rm d}\phi^2\\[4pt]
& = &\left(1+\frac{\omega^2\rho^2\lambda'\,^2}{1+\omega^2\rho^2}\right){\rm d}\rho^2+(1+\omega^2\rho^2)\left({\rm d}\phi+\frac{\lambda'\,^2}{1+\omega^2\rho^2}{\rm d}\rho\right)^2\,,
\end{eqnarray}
one finds the desired coordinate system $(\xi,\chi)=(\xi(\rho,\phi),\chi(\rho,\phi))$ by solving
\begin{equation}
\left\{
\begin{array}{ccc}
{\rm d}\xi & = & \displaystyle\left(1+\frac{\omega^2\rho^2\lambda'\,^2}{1+\omega^2\rho^2}\right)^{1/2}{\rm d}\rho\\ [10pt]
{\rm d}\chi & = & \displaystyle\frac{\lambda'}{1+\omega^2\rho^2}\,{\rm d}\rho+{\rm d}\phi
\end{array}
\right.\,.
\end{equation}
Observe that our notations are slightly distinct from that of Do Carmo and Dajczer \cite{DoCarmoTohoku1982}: $(\xi,\chi,\omega\phi,\omega,a)_{ours}\mapsto(s,t,\phi,1/h,m)_{theirs}$. 

Using natural parameters $(\xi,\chi)$ to write the line element gives
\begin{equation}
{\rm d}s^2 = {\rm d}\xi^2+(1+\omega^2\rho^2){\rm d}\chi^2,
\end{equation}
which, by taking into account that $\rho$ does not depends on $\chi$, i.e., $\rho=\rho(\xi)$, and consequently also $\lambda=\lambda(\xi)$, can be rewritten as
\begin{equation}
{\rm d}s^2 = {\rm d}\xi^2+\mathcal{U}^2(\xi){\rm d}\chi^2,
\end{equation}
where $\mathcal{U}^2(\xi)=1+\omega^2\rho^2(\xi)$. For a helicoid, the map  $(\rho,\phi)\mapsto(\xi,\chi)$ is just the identity and, therefore, one has $\mathcal{U}^2_{helicoid}(\rho=\xi)=1+\omega^2\rho^2$. 

The function $\mathcal{U}$ encodes all the geometric information of its associated helicoidal surface and, consequently, both the Gaussian and the Mean curvatures are written in terms of $\mathcal{U}$. Further, we mention that $\mathcal{U}$ also determines the geometry-induced potential which governs the behavior of a quantum particle confined on the associated helicoidal surface.
\newline

A natural question now is if we can associate a helicoidal surface with a given non-negative function $\tilde{\mathcal{U}}(\xi)$, i.e., 
\newline
{\it Problem:} Given a function $\tilde{\mathcal{U}}(\xi)>0$, is it possible to find a constant $\tilde{\omega}$ and some functions $\rho,\phi$, and $\lambda$, such that the helicoidal surface $$X(\rho,\phi)=(\rho\cos(\tilde{\omega}\phi),\rho\sin(\tilde{\omega}\phi),\lambda(\rho)+\phi)$$ has its line element written in natural coordinates as ${\rm d}s^2={\rm d}\xi^2+\tilde{\mathcal{U}}^2\,{\rm d}\chi^2$?

This problem do admit a solution to any given function $\mathcal{U}>0$. In fact, it is always possible to find a 2-parameter family of helicoidal surfaces associated with it. This is precisely the content of the {\it Bour Lemma} \cite{DoCarmoTohoku1982}. It states that for every non-zero function $\mathcal{U}$ there exists a 2-parameter family of isometric helicoidal surfaces associate with it. The functions $(\rho,\phi)$ and $\lambda(\rho)$ which characterize the helicoidal surface can be written as \cite{DoCarmoTohoku1982}
\begin{equation}
\left\{
\begin{array}{l}
\rho=\rho(\xi)=\displaystyle\frac{1}{\omega}\sqrt{a^2\,\mathcal{U}^2-1}\\[10pt]
\lambda=\lambda(\xi)=\displaystyle\frac{1}{\omega}\int {\rm d}\xi\,\frac{a\,\mathcal{U}}{a^2\,\mathcal{U}^2-1}\sqrt{a^2\mathcal{U}^2\,[\omega^2-a^2\,\dot{\mathcal{U}}^2]-\omega^2}\\[8pt]
\phi=\phi(\xi,\chi)=\displaystyle\frac{\chi}{a}-\frac{1}{\omega}\displaystyle\int {\rm d}\xi\,\frac{\sqrt{a^2\,\mathcal{U}^2\,[\omega^2-a^2\,\dot{\mathcal{U}}^2]-\omega^2}}{a\,\mathcal{U}[a^2\,\mathcal{U}^2-1]}
\end{array}
\right.,\label{Eq::solucaoLemaBour}
\end{equation}
where a dot represent the derivative with respect to $\xi$: $\dot{\mathcal{U}}={\rm d}\,\mathcal{U}/{\rm d}\xi$. By varying the constants $a$ and $\omega$ above, we generate a 2-parameter family of isometric helicoidal surfaces associated with the $\mathcal{U}$ given {\it a priori}\footnote{If we choose $\mathcal{U}=\mathcal{U}_0$ to be a constant function, then we obtain a 2-parameter family of helicoidal surfaces which are contained on a cylinder of radius $\rho=\sqrt{a^2\mathcal{U}^2_0-1}$.}.

Finally, the Gaussian and Mean curvatures are written as \cite{DoCarmoTohoku1982}
\begin{equation}
K=K(\xi) = - \frac{\ddot{\mathcal{U}}}{\mathcal{U}},\label{eqGaussCurvHelicoidalSurfaces} 
\end{equation}
and
\begin{equation}
H=H(\xi)=\frac{a^2\, \mathcal{U}\,\ddot{\mathcal{U}}+a^2\,\dot{\mathcal{U}}^2-\omega^2}{2\sqrt{a^2\,\mathcal{U}^2\,[\omega^2-a^2\,\dot{\mathcal{U}}^2]-\omega^2}}\label{eqMeanCurvHelicoidalSurfaces}
\end{equation}
respectively, where he have adopted the surface normal $\mathbf{N}=\mathcal{U}^{-1}(\partial_{\chi}X\times\partial_{\xi}X)$.

According to the Bour lemma, we have for each function $\mathcal{U}$ a 2-parameter family of isometric helicoidal surfaces $[\mathcal{U},\omega,a]$. This means that the metric, and also the Gaussian curvature, is the same for all the helicoidal surfaces in the family. However, since the Mean curvature is not a bending invariant, the parameters $\omega$ and $a$ can give rise to different values of $H$. It follows that these parameters can be of physical relevance, since the geometry-induced potential also depends on the Mean curvature $H$.
\newline
\newline
{\it Example 6.1:} (helicoidal minimal surfaces) Imposing the condition $H=0$ to Eq. (\ref{eqMeanCurvHelicoidalSurfaces})
gives
\begin{eqnarray}
a^2\mathcal{U}\,\ddot{\mathcal{U}}+a^2\dot{\mathcal{U}}^2=\omega^2\,\Rightarrow\,\mathcal{U}^2(\xi)=\frac{1}{a^2}(\omega^2\xi^2+2\,\omega_1\omega\xi+\omega_0),
\end{eqnarray}
where $\omega_0$, $\omega_1$ are constants  satisfying $b=\omega_0-\omega_1^2\geq1$, since $a^2\mathcal{U}^2-1>0$. In short, helicoidal minimal surfaces are characterized by a quadratic polynomial (for the particular case of a helicoid, we have $a=\omega_0=1$ and $\omega_1=0$). The Gaussian curvature of a helicoidal minimal surface is given by
\begin{equation}
K(\xi)=-\frac{\omega^2(\omega_0-\omega_1^2)}{a^4\mathcal{U}^4}=-\frac{b\,\omega^2}{[(\omega\xi+\omega_1)^2+b]^2}<0\,.\label{eqGaussianCurvMinHelSurf}
\end{equation}

The solution of Eq. (\ref{Eq::solucaoLemaBour}) for a helicoidal minimal surface is
\begin{equation}
\left\{
\begin{array}{l}
\rho=\rho(\xi)=\displaystyle\frac{1}{\omega}\sqrt{\omega^2\xi^2+2\,\omega_1\omega\xi+\omega_0-1}\\[8pt]
\lambda(\xi)=\sqrt{b-1}\displaystyle\int {\rm d}\xi\,\frac{1}{\sqrt{\omega^2\xi^2+2\,\omega_1\omega\xi+\omega_0}\,(\omega^2\xi^2+2\,\omega_1\omega\xi+\omega_0-1)}\\[10pt]
\phi=\phi(\xi,\chi)=\displaystyle\frac{\chi}{a}-\sqrt{b-1}\displaystyle\int {\rm d}\xi\,\frac{\sqrt{\omega^2\xi^2+2\,\omega_1\omega\xi+\omega_0}}{\omega^2\xi^2+2\,\omega_1\omega\xi+\omega_0-1}
\end{array}
\right..
\end{equation}
The parameter $a$ plays no relevant role. Indeed, by doing $\chi\mapsto a\chi$, we see that all the surfaces with distinct $a$ have the same image but different parametrizations. 

\section{Schr\"odinger equation on invariant surfaces}

In the previous sections, we have introduced and studied the geometry of surfaces invariant by a 1-parameter subgroup of isometries of $\mathbb{R}^3$. In this section, we devote our attention to the Schr\"odinger equation for a constrained particle on such surfaces.

All the three types of invariant surfaces have in common the following property: they admit the existence of a coordinate system $(u,v)$ such that the respective line element can be written as
\begin{equation}
{\rm d}s^2 = {\rm d}u^2 + f^2(u)\,{\rm d}v^2,\,(u,v)\in[u_0,u_1]\times[v_0,v_1],
\end{equation}
where $f$ is a positive smooth function. Since such a metric has $g_{12}=0$, the Gaussian curvature (which only depends on the coefficients $g_{ij}$) can be expressed as
\begin{equation}
K = -\frac{1}{2\sqrt{g_{11}g_{22}}}\left[\frac{\partial}{\partial v}\left(\frac{g_{11,2}}{\sqrt{g_{11}g_{22}}}\right)+\frac{\partial}{\partial u}\left(\frac{g_{22,1}}{\sqrt{g_{11}g_{22}}}\right)\right] = -\frac{\ddot{f}}{f},
\end{equation}
where $g_{ij,k}=\partial g_{ij}/\partial q^k$, with $q^1=u$, $q^2=v$, and a dot denotes the derivative with respect to $u$. Naturally, the Mean curvature does not admit such a unified description, since it is not a bending invariant.

Let $\Sigma$ be an invariant surface with coordinate system $(u,v)$ as above. The Hamiltonian reads
\begin{equation}
\hat{H} = -\frac{\hbar^2}{2m}\Delta_g+V_{gip}=-\frac{\hbar^2}{2mf}\left[\frac{\partial}{\partial u}\left(f\frac{\partial}{\partial u}\right)+\frac{1}{f}\frac{\partial^2}{\partial v^2}\right]+V_{gip}\,.
\end{equation}
Now, rescaling the wave function as $\psi \mapsto \psi/g^{1/4}=\psi/\sqrt{f}$ (the Hamiltonian $\hat{H}$ should be rescaled as $f^{\frac{1}{2}}\,\hat{H}\,f^{-\frac{1}{2}}$), we have
\begin{eqnarray}
\hat{H} & = & -\frac{\hbar^2}{2m}\left[\frac{\partial^2}{\partial u^2}+\frac{1}{f^2}\frac{\partial^2}{\partial v^2}\right]+V_{eff}\,,
\end{eqnarray}
where
\begin{eqnarray}
V_{eff} &=&-\frac{\hbar^2}{2m}\left(-\frac{\ddot{f}}{2f}+\frac{\dot{f}\,^2}{4f^2}\right)-\frac{\hbar^2}{2m}(H^2-K),\\
& = & -\frac{\hbar^2}{2m}\left(\frac{\dot{f}\,^2}{4f^2}+\frac{\ddot{f}}{2f}\right)-\frac{\hbar^2}{2m}\,H^2\,.
\end{eqnarray}

As a corollary, it follows that the stationary Schr\"odinger equation can be solved by separation of variables. Indeed, writing $\psi(u,v)=A(u)B(v)$, we have
\begin{equation}
\frac{1}{B(v)}\frac{{\rm d}^2B(v)}{{\rm d}v^2}=-\Big(U(u)+k^2\Big)f^2(u)-\frac{f^2(u)}{A(u)}\frac{{\rm d}^2A(u)}{{\rm d}u^2},
\end{equation}
where  $k^2=2mE/\hbar^2$ and $U(u)=-2m\,V_{eff}(u)/\hbar^2$. This procedure   furnishes the following equations
\begin{equation}
\left\{
\begin{array}{c}
B''(v) =-\lambda B(v) \\[5pt]
A''(u) + \left(U(u)+k^2-\displaystyle\frac{\lambda}{f^2(u)}\right)A(u) = 0\\
\end{array}
\right.\,,\label{eq::EqSchrDesacopladaSuperInvar}
\end{equation}
whose solutions depends on the imposed boundary conditions. 

The above equations clearly show that for an invariant surface the stationary Schr\"odinger equation decouples into an equation along the orbits of the 1-parameter subgroup ($v$-curves) and an effective equation along the direction orthogonal to the orbits ($u$-curves), i.e., an effective equation along the generating curve.

\subsection{Schr\"odinger equation for cylindrical surfaces}

For a cylindrical surface one has 
\begin{equation}
\left\{\begin{array}{c}
{\rm d}s_{cyl}^2 = {\rm d}u^2+{\rm d}v^2\\[5pt]
K_{cyl}\equiv0,\,H_{cyl} = \displaystyle\frac{\dot{x}\ddot{y}-\ddot{x}\dot{y}}{2} = \frac{\kappa}{2}\\
\end{array}
\right.,
\end{equation}
where $\kappa(u)$ is the curvature function of the cross section $\alpha(u)=(x(u),y(u),0)$ (generating curve), $u$ being its arc-length, which is translated in the direction of $(0,0,1)$. Thus, the decoupled equations (\ref{eq::EqSchrDesacopladaSuperInvar}) read
\begin{equation}
\left\{
\begin{array}{c}
B''(v) =-\lambda B(v)\\[5pt]
A''(u) +\frac{\kappa^2}{8}A(u)+(k^2-\lambda)A(u) = 0\\
\end{array}
\right.\,.\label{eq::EqSchrDesacopladaSuperCylind}
\end{equation}
For a cylindrical surface we may assume homogeneous boundary conditions for the $v$-directions. Then, the energy spectrum is given by
\begin{equation}
E_{cyl}(n_u,n_v) = \frac{h^2n_v^2}{8mL_v^2} + E_{\kappa,n_u}\,,
\end{equation}
where $L_v$ is the height of the cylindrical surface, with $n_v\in\{1,2,...\}$, and $E_{\kappa,n_u}$ is the $n_u$-th eigenenergy of a constrained particle in a 1D box of length $L_u$ under a potential $V_{gip}=-\hbar^2\kappa^2/8m$: a box with homogeneous or periodic boundary conditions if $\alpha$ is open or closed, respectively.\footnote{In an intrinsic approach, i.e., in the absence of $V_{gip}=-\hbar^2\kappa^2/8m$, one would find $E_{cyl}(n_u,n_v) = h^2n_v^2/8mL_v^2 + h^2n_u^2/8mL_u^2$, with $n_u,n_v\in\{1,2,...\}$, for an open cross section or $E_{cyl}(n_u,n_v) = h^2n_v^2/8mL_v^2 + h^2n_u^2/2mL_u^2$, with $n_u\in\{1,2,...\}$ and $n_v\in\{0,1,2,...\}$, for a closed cross section \cite{BastosEtAlPreprint}.}.

\subsection{Schr\"odinger equation for surfaces of Revolution}

For a surface of revolution one has 
\begin{equation}
\left\{\begin{array}{c}
{\rm d}s_{rev}^2 = {\rm d}u^2+x^2(u){\rm d}v^2\\[5pt]
K_{rev}=-\frac{\ddot{x}}{x},\,H_{rev} = \displaystyle\frac{x(\dot{x}\ddot{z}-\ddot{x}\dot{z})+\dot{z}}{2x}\\
\end{array}
\right.,
\end{equation}
where $\alpha(u)=(x(u),0,z(u))$, with $x>0$, is the generating curve which is rotated around the $z$ axis, with arc-length parameter $u$. Then, the decoupled equations (\ref{eq::EqSchrDesacopladaSuperInvar}) read
\begin{equation}
\left\{
\begin{array}{c}
B''(v) =-\lambda B(v)\\[5pt]
A''(u) +\left(\displaystyle\frac{\dot{x}^2}{4x^2}+\frac{\ddot{x}}{2x}+H_{rev}^2\right)A(u)+ \left(k^2-\displaystyle\frac{\lambda}{x^2}\right)A(u) = 0\\
\end{array}
\right.\,.\label{eq::EqSchrDesacopladaSuperRevol}
\end{equation}
For a revolution surface we may assume an angular periodicity for the $v$-curves, which gives $\lambda=m_{\chi}^2$, $m_{\chi}\in\mathbb{Z}$. Then, the effective dynamics in the $u$-direction is
\begin{equation}
-\frac{\hbar^2}{2m}A''(u) -\frac{\hbar^2}{2m}\left(\displaystyle\frac{\dot{x}^2+2x\ddot{x}-4m_{\chi}^2}{4x^2}+H_{rev}^2\right)A(u) = E\,A(u)\,.
\end{equation}
Depending on the concavity of $x(u)$ and on the values of the angular momentum quantum number $m_{\chi}$, the contribution of $\dot{x}^2+2x\ddot{x}-4m_{\chi}^2$ in the effective potential for the $u$-direction can be attractive or repulsive, then changing the way it favors the existence of geometry-induced bound states \cite{MoraesAnnPhys,AtanasovPLA2007}.

\subsection{Schr\"odinger equation for helicoidal surfaces}

For a helicoidal surface one has 
\begin{equation}
\left\{\begin{array}{c}
{\rm d}s_{hel}^2 = {\rm d}u^2+\mathcal{U}^2(u){\rm d}v^2\\[5pt]
K_{hel}=-\displaystyle\frac{\ddot{\mathcal{U}}}{\mathcal{U}},\,H_{hel}=\displaystyle\frac{a^2\, \mathcal{U}\,\ddot{\mathcal{U}}+a^2\,\dot{\mathcal{U}}^2-\omega^2}{2\sqrt{a^2\,\mathcal{U}^2\,[\omega^2-a^2\,\dot{\mathcal{U}}^2]-\omega^2}}\\
\end{array}
\right.,
\end{equation}
where $u=\xi$ and $v=\chi$ are natural parameters of the helicoidal surface introduced in section 6. Then, the decoupled equations (\ref{eq::EqSchrDesacopladaSuperInvar}) read
\begin{equation}
\left\{
\begin{array}{c}
B''(v) =-\lambda B(v)\\[5pt]
A''(u) +\left(\displaystyle\frac{\dot{\mathcal{U}}^2}{4\,\mathcal{U}^2}+\frac{\ddot{\mathcal{U}}}{2\,\mathcal{U}}+H_{hel}^2\right)A(u)+ \left(k^2-\displaystyle\frac{\lambda}{\mathcal{U}^2}\right)A(u) = 0\\
\end{array}
\right.\,.\label{eq::EqSchrDesacopladaSuperRevol}
\end{equation}

The standard example of a helicoidal surface is that of a helicoid. For such a surface it is known that particles with distinct angular quantum numbers tend to localize in distinct parts of the helicoid and also that there exist geometry-induced bound states \cite{AtanasovPRB2009}. In the following we extend these findings to all helicoidal minimal surfaces (the helicoid being the simplest example) and, due to the existence of other parameters associated to a helicoidal minimal surface, we show in addition the possibility of controlling the change in the distribution of the probability density when the surface is subjected to an extra charge, i.e., where the particles are find with greatest probability.

\subsection{Constrained dynamics on helicoidal minimal surfaces} 

\begin{figure*}[tbp]
\centering
  {\includegraphics[width=0.45\linewidth]{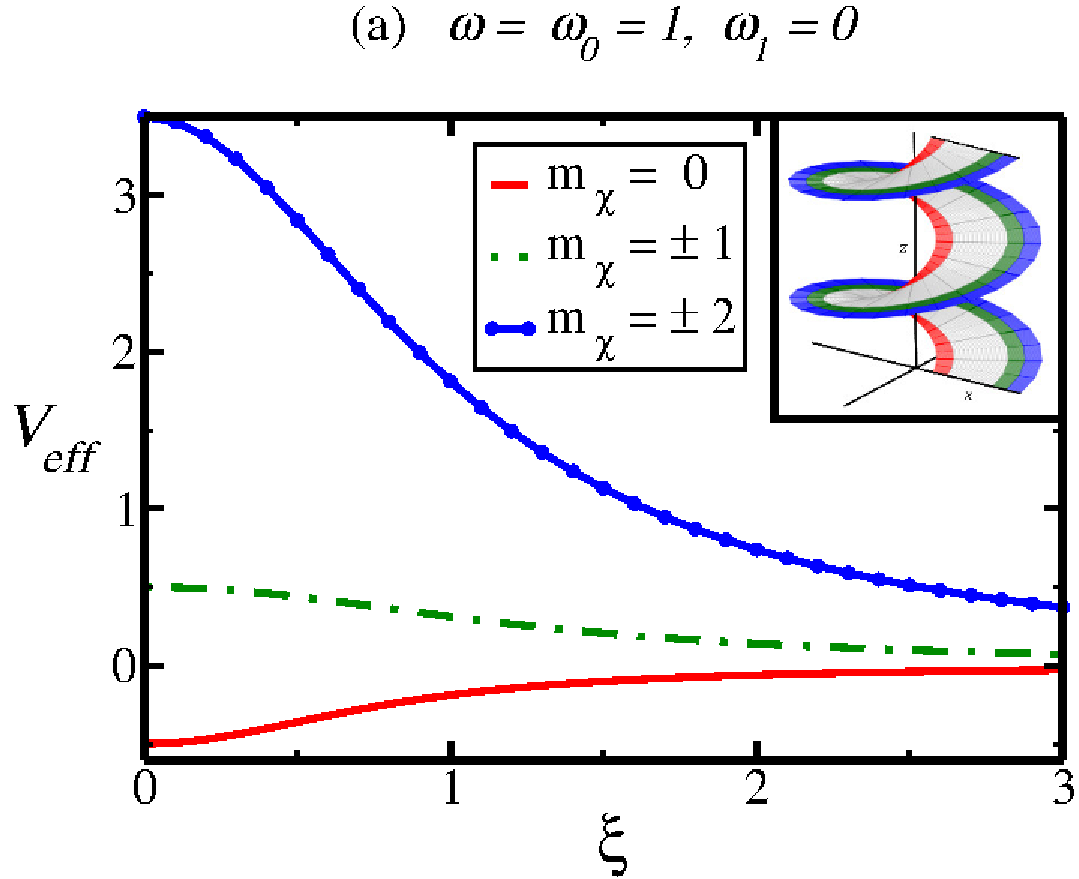}  {\includegraphics[width=0.45\linewidth]{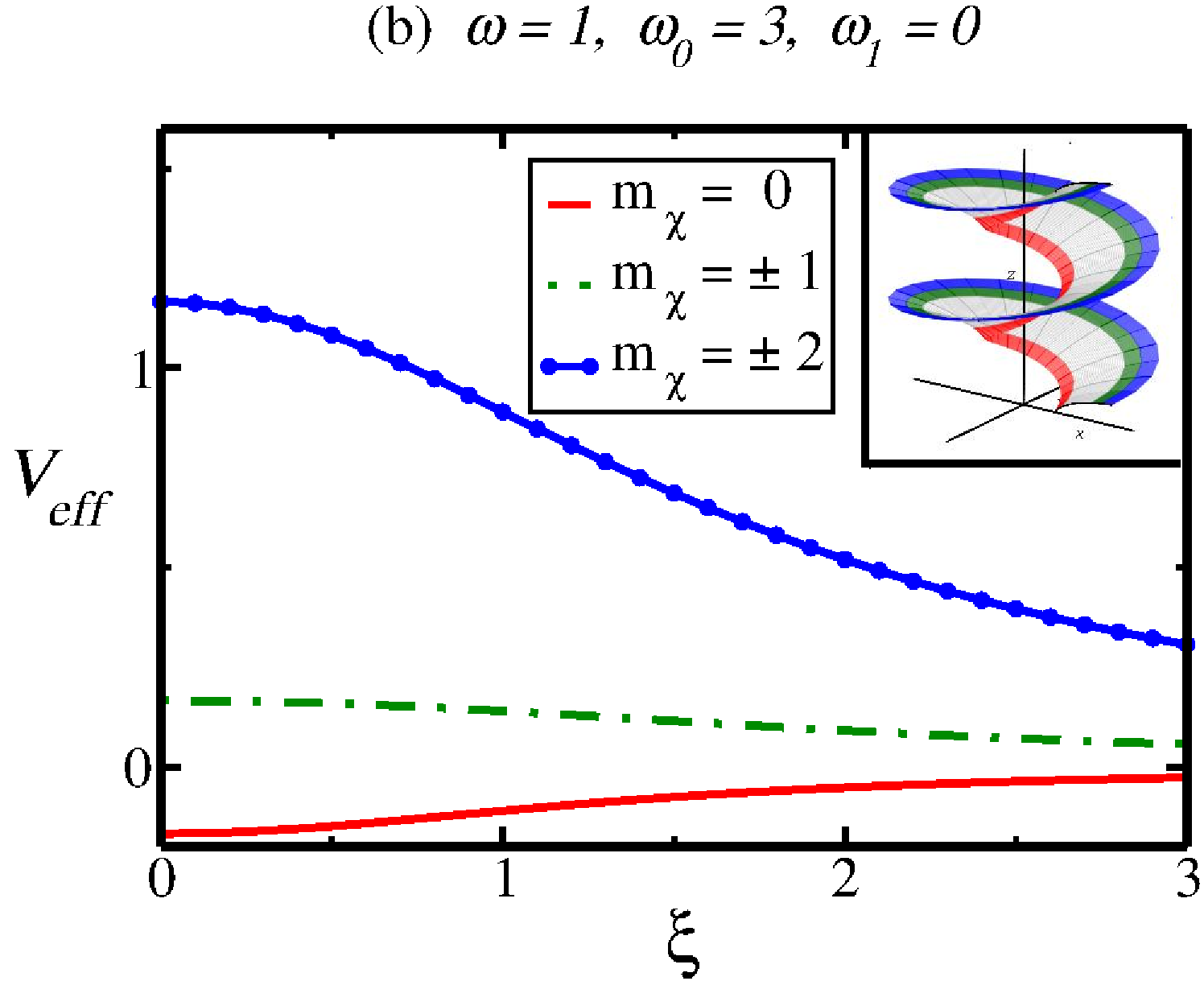}}}
  
          \caption{The behavior of the effective potential $V_{eff}$ as a function of $\xi$ with $\hbar^2/2m=1$, $\omega=1$, $\omega_1=0$, and different values of $m_{\chi}$. Insets:  helicoidal surfaces for (a) $\omega_0=1$ and (b) $\omega_0=3$, respectively. The insets illustrate the fact that particles with distinct values of $m_{\chi}$ tend to localize in different parts of the surface.
          }
\label{fig:Fig1}
\end{figure*}

For a helicoidal minimal surface one has $H\equiv0$ and $\mathcal{U}^2=(\omega\xi+\omega_1)^2+b$ as seen in Example 6.1 (without loss of generality, we set $a=1$). Then, the decoupled Schr\"odinger equation reads ($u=\xi$, $v=\chi$)
\begin{equation}
B''(\chi) =-\lambda B(\chi)
\end{equation}
and
\begin{equation}
A''(\xi) + \frac{\omega^2}{4}\left\{\displaystyle\frac{1-\lambda}{b+(\omega\xi+\omega_1)^2}+\frac{b}{[b+(\omega\xi+\omega_1)^2]^2}\right\}A(\xi)
+k^2A(\xi) = 0\,.\label{eqfuncAforMHS}
\end{equation}

Writing the solution for $B$ as $B(\chi)={\rm e}^{{\rm i}k_{\chi}\,\chi}$ furnishes
\begin{equation}
\lambda = k_{\chi}^2,
\end{equation}
with $k_{\chi}$ being the partial moment in the $\chi$ direction. The canonical momentum associated to the coordinate $\chi$, $L_{\chi}=-{\rm i}\,\hbar/\omega\,\partial_{\chi}$, has the same eigenfunctions as the equation for $B$. The momentum $k_{\chi}$ is quantized according to 
\begin{equation}
k_{\chi} = m_{\chi}\,\omega\,,m_{\chi}\in\mathbb{Z}\,.
\end{equation}

Using the expression for $\lambda$ in Eq. (\ref{eqfuncAforMHS}) shows that the equation for the $\xi$ direction is subjected to the following effective potential
\begin{equation}\label{Veff}
V_{eff}(\xi)=-\frac{\hbar^2}{2m}\frac{\omega^2}{4}\left\{\frac{b}{[b+(\omega\xi+\omega_1)^2]^2}+\displaystyle\frac{1-4\,m_{\chi}^2}{b+(\omega\xi+\omega_1)^2}\right\}\,.
\end{equation}

This effective potential displays two terms with distinct contributions. The first term contributes attractively, while the second one depends on the sign of $m_{\chi}^2-1/4$, acting attractively for $m_{\chi}^2<1/4$, i.e., $m_{\chi}=0$, or repulsively for $m_{\chi}^2>1/4$, i.e., $m_{\chi}\not=0$. The effect of this variable part is of a centrifugal potential character for $m_{\chi}\not=0$ (repulsive), it pushes a particle to the outer border of the surface. On the other hand, when $m_{\chi}=0$ (attractive), the contribution of this variable part is of a anticentrifugal character and it concentrates the particles in the inner border of the minimal helicoidal surface, i.e., around the screw axis. This analysis is in agreement with what happens for the particular case of a helicoid \cite{AtanasovPRB2009}, where $\omega_0=1$ and $\omega_1=0$ ($b=1$). 

Now, we benchmark our analytical expression for the effective potential [see Eq. (\ref{Veff})] with the one derived by Atanasov {\it et al.} \cite{AtanasovPRB2009}. For that purpose, we can assume the following set of parameters: $\omega_1 =0$ and $\omega_0 =1$ (then $b=1$). After substitution of values, we find that
\begin{equation}\label{VeffAtanosov}
V_{eff}(\xi)=-\frac{\hbar^2}{2m}\frac{\omega^2}{4}\left[\frac{1}{(1+\omega^2\xi^2)^2}+\displaystyle\frac{1-4\,m_{\chi}^2}{1+\omega^2\xi^2}\right]\,.
\end{equation}
The evolution of $V_{eff}$, in Eq. (\ref{VeffAtanosov}), as a function of $\xi$ is depicted in Fig. \ref{fig:Fig1} (a). As one readily sees, the behavior of $V_{eff}$ is strongly affected by the angular momentum quantum number $m_{\chi}$. When
$m_{\chi} =0$, we have $V_{eff}(\xi) < 0$, which leads to the existence of bound states\footnote{A globally attractive potential $V$ satisfying the criterion $\int V(x)\,{\rm d}x^n<0$ do admit the existence of bound states for $n=1$ or $n=2$ \cite{Chadan2003}.}. On the other hand, for nonvanishing angular momentum quantum numbers, we observe that $V_{eff}(\xi) > 0$ (the energy spectrum is positively valued) and no bound state is allowed \cite{AtanasovPRB2009}.

The above analysis is still valid for other values of the parameters $\omega$, $\omega_1$, and $\omega_0$. In other words, the existence of geometry-induced bound and localized states previously verified for a helicoid \cite{AtanasovPRB2009} can be extended to any helicoidal minimal surface.
\newline

Finally, let us comment that other results established for a helicoid can be extended to all helicoidal minimal surfaces with some additional advantages. Indeed, applying a change of variables
\begin{equation}
\xi(\tilde{\xi})=\sqrt{b}\,\,\tilde{\xi}-\frac{\omega_1}{\omega}=\sqrt{(\omega_0-\omega_1^2)}\,\,\tilde{\xi}-\frac{\omega_1}{\omega},\label{eq::MappingFromHelMinSurfToAhelicoid}
\end{equation}
which implies ${\rm d}A/{\rm d}\tilde{\xi}=\sqrt{b}\,{\rm d}A/{\rm d}\xi$, we can map our effective equation for the $\xi$-direction into that of a helicoid \cite{AtanasovPRB2009}
\begin{equation}
-\frac{\hbar^2}{2m}\frac{{\rm d}^2A}{{\rm d}\,\tilde{\xi}^2}-\frac{\hbar^2}{2m}\frac{\omega^2}{4}\left\{\frac{1}{(1+\omega\tilde{\xi} ^2)^2}+\displaystyle\frac{1-4\,m_{\chi}^2}{1+\omega\tilde{\xi}^2}\right\}A(\tilde{\xi}) = b\,E\,A(\tilde{\xi})\,.
\end{equation}
For example, when analyzing the distribution of the probability density for a constrained particle on a helicoidal minimal surface subjected to some charge distribution, as analyzed by Atanasov {\it et al.} \cite{AtanasovPRB2009}, we can use the correspondence above to map the problem for a helicoidal minimal surface into an equivalent problem for a helicoid and then, by inverting Eq. (\ref{eq::MappingFromHelMinSurfToAhelicoid}), solve the original problem. So, by tuning the parameters $\omega_0$ and $\omega_1$, we can control the changing in the distribution of the probability density, e.g., we can govern the location where the particle will be found with greatest probability when the outer border of the helicoidal minimal surface is uniformly charged (i.e., where the extra charge will concentrate).

\section{\label{sec:concl}Conclusions} 

The main goal of this work was the study of the problem of prescribed geometry-induced potential for curves and surfaces in the 3D Euclidean space, i.e., how to find a curve or a surface with a potential given {\it a priori}, whose solution offers the possibility of engineering surfaces and curves with a quantum behavior prescribed {\it a priori} through their geometry-induced potential.

It is shown that the prescribed potential problem for curves can be solved by integrating the Frenet equations, whose solutions can be explicitly found for planar curves, while the analogous problem for surfaces involves the solution of a non-linear 2nd order PDE. We further restricted ourselves to the study of surfaces invariant by a 1-parameter group of isometries of $\mathbb{R}^3$, which led us to the study of cylindrical, revolution, and helicoidal surfaces. Due to their appealing symmetry, i.e. translation, rotation, and screw symmetry, these surfaces are commonly encountered in applications and theoretical studies of quantum mechanics and do not constitute any severe restriction to the investigation of a constrained dynamics on surfaces. Besides, this simplifying hypothesis turns the study of the PDE for the prescribed potential into that of an ODE and discloses many potentialities of invariant surfaces in applications. In addition, the invariance property also allows for a unified description of the Schr\"odinger equation under the effect of a geometry-induced potential. We completely solved the problem for cylindrical and revolution surfaces. For the class of helicoidal surfaces we presented the concept of natural parameters, which allows for a unified description of such surfaces and also the association of a 2-parameter family of isometric helicoidal surfaces with a prescribed positive function. These surfaces are  particularly important due to the fact that, by screw-rotating a curve clockwisely and counterclockwisely, one can easily generate pairs of enantiomorphic surfaces, which naturally turns helicoidal surfaces an adequate setting to test and exploit a link between chirality and the effects of a geometry-induced potential. Finally, for the family of helicoidal minimal surfaces we proved the existence of geometry-induced bound and localized states, then generalizing known results for the particular case of a helicoid, and in addition we also showed the possibility of controlling the change in the distribution of the probability density when the surface is subjected to an extra charge.

Naturally, for a more realistic description of the constrained dynamics, the approach presented here must be
extended to the context of a relativistic dynamics, such as the Dirac equation. This is presently under investigation and will be the subject of a follow-up work.

\section*{Acknowledgments}

The authors would like to thank useful discussions with F. A. N. Santos, F. Moraes, R. T. Gomes, G. G. Carvalho, and J. Deibsom da Silva, during the preparation of the manuscript, and also thank the financial support by CNPq, CAPES, and FACEPE (Brazilian agencies).

\section*{References}


\providecommand{\noopsort}[1]{}\providecommand{\singleletter}[1]{#1}%

\end{document}